\documentstyle[preprint,aps]{revtex}

\begin{document}
\title{ac Susceptibility in Granular Superconductors: Theory and Experiment.}
\author{A. Kunold, M. Hern\'{a}ndez, A. Myszkowski, J.L. Cardoso, P. Pereyra}
\address{Departamento de Ciencias B\'asicas, UAM/Azcapotzalco, Av. S. Pablo 180,\\
M\'exico D.F. 02200}
\date{\today}
\maketitle

\begin{abstract}
A phenomenological theory to describe the electromagnetic properties of
granular superconductors, based on known bulk superconductors expressions
and conventional Josephson's junctions tunneling currents, is presented and
successfully used to fit distinct experimental results for the magnetic
susceptibility $\chi $ as a function of the temperature and the applied
magnetic field of rather different samples.
\end{abstract}

\section{Introduction}

The influence of the granular structure on the electromagnetic
superconducting properties has been consistently recognized \cite{Clem}, and
simple formulas and models were used to account for the main features of the
ac susceptibility $\chi (B,T)$ as a function of temperature $(T)$ and
magnetic field $(B)$. The critical state model\cite{Bean}, and its extended
versions\cite{Clem,Kimishima}, have been useful in interpreting the observed
values of the magnetic response of type-II superconductors and the
underlying interplay between pinning and creep forces on Josephson vortices
in the grain and intergrain regions, both in zero-field cooling and field
cooling experiments \cite{Murphy,Kogan,Beadoin,Bruneel,Ravi,de Lima}. Other
effects, such as phase coherence and decoherence phenomena, peculiar to
Josephson junctions and the quantum interference effects in the loops, add
extra complexity to the polycrystalline magnetic response. In general, these
effects have been neglected or taken into account only partially (or
indirectly) to describe the ac susceptibility in polycrystalline
superconductors. It is the purpose of this paper to consider them
explicitly, to deduce a simple formula to describe the granular
superconductors susceptibility in terms of the Josephson junction and loop
areas, the penetration depth, the decoherence decay length and a critical
exponent, and to show the ability of the formula to account qualitatively
and quantitatively the rather complex behavior of $\chi (B,T)$. To
illustrate this, magnetic response data of quite different samples are
considered and fitted with fairly good results.

The understanding of the granular structure influence on the electromagnetic
properties is not only of relevant interest but also of a challenging
nature. Many effects, related with the phase coherence and Josephson
vortices pinning and displacements, combine to give rise to fluorishly
different experimental results. Careful measurements reveal subtle and
systematic quantum coherence phenomena that affect the order parameter of
adjacent superconducting grains and produce well defined field- and
sample-dependent oscillations, apparent in the critical transport and
screening currents behavior\cite
{discorto,gaps,imchi,paramag,anisotro,ceramic,dislargo}.

In the last years we have been interested in high precision measurements of
critical transport and screening currents as functions of the temperature,
magnetic field and the granularity of the superconducting samples. The main
idea in these experiments was to reduce the effects of macroscopic averaging 
\cite{ourspb,ourspc1,ourspc2}. Recently, a phenomenological theory of the
Josephson junction between two superconducting grains, taking into account
the effects of coherence and decoherece decay lengths on the order
parameter, has been suggested to describe the Ambegaokar-Baratoff to
Ginsburg-Landau transition in the temperature behavior of the critical
transport current \cite{pepe1}.

In the present paper, we study the behavior of the magnetic susceptibility $%
\chi $ as a function of the temperature and the external magnetic field. The
measurement of $\chi $ permits to study the tunneling phenomena far away
from the critical points and limits the current paths to a few adjacent
grains. For this report we consider two quite different samples: a ''bad''
and porous sample B, with irregular grains and low magnetic susceptibility,
and a ''good'' sample G, with closely packed parallelepiped shaped grains
and much higher value of $\chi $ (see Figure 1). The ''bad'' sample
susceptibility also presents oscillations and a pronounced decrease of $\chi 
$ with increasing magnetic field (see Figure 2).

The contentsof this paper are as follows. In the second Section we recast
the Clem's picture of the screening currents in granular systems. We then
recall, in Section III, some basic and known expressions of the
superconducting theory and `put them together' in a kind of phenomenological
formula to describe the real part of the ac susceptibility. In Section IV,
this phenomenological theory is applied to rather different granular
superconductors, and finally in Section V we give a brief discussion.

\section{Screening currents in granular systems}

It is well known that the basic magnetic properties of ceramics are created
by the currents circulating inside the grains, which tend to expel the
external magnetic field from parts of the volume (regions of the type I in
Figure 3) and thus lead to a diamagnetic behavior. Following Clem's picture
of the granular behavior, we distinguish three characteristic regions as
shown in Figure 3. In this Figure, the screening current in regions II,
responsible for the basic diamagnetic properties of the material, is
distributed as suggested in the lower part of the Figure. The current
density reaches its maximum (critical value) at the boundary of this region
II, and beyond this region's ''surface'', i.e. in region III , there is no
screening current.

Josephson tunneling junctions are formed in regions III and IV and the
Josephson tunneling process connects the `surfaces' of regions II in
adjacent grains, which are of course in a critical state. It is important to
stress that, even if the applied magnetic field is weak, far away from its
critical value (from the macroscopic point of view), the screening currents
and the local magnetic field on each grain determine a local criticality and
the penetration length, the existence or non existence of screening currents
and, consequently, the position of the surfaces limiting the regions II,
which (from the microscopical point of view) are in critical condition.

The ceramic structure permits the flowing of circulating currents over
clusters of grains. The represented cluster of three grains (in Figure 4) is
only an example; there is a possibility of current paths over a great amount
of grains with Josephson type junctions between each pair. The corresponding
currents produce some additional magnetic effects and modify the magnitude
of $\chi $. These currents are, of course, regulated by the quantum
interference phenomena and the Josephson junctions.

In Figure 4, the current 1 flowing over the external part of a cluster
circulates in the same direction as the intragranular currents. This current
has a diamagnetic character and raises the value of $\left| \chi \right| $.
However, it does not expel the external magnetic field completely from the
junctions. The internal part of the circuit contains the current 2, flowing
in the opposite direction, though alongwith the currents inside the grains,
it is of paramagnetic character\cite{paramag,paramag2,hein}, and permits the
passage of some vortices of external magnetic field through the
intergranular region V.

\section{Theory of the magnetic response in polycrystalline samples}

As in the case of the critical transport current, the behavior of the
magnetization current presents systematic sample-dependent oscillations.
These oscillations in polycrystalline and highly random systems, with low
density of irregular grains, can also be explained reasonably well in terms
of the superposition of Josephson junction and quantum-interfering
intergrain currents. Therefore, the total current contains the contributions
of the Josephson type currents 
\begin{equation}
j_{J}=j_{c}\frac{\sin (\pi \phi _{J}/\phi _{o})}{\pi \phi _{u}/\phi _{o}}
\end{equation}
and loop currents 
\begin{equation}
j_{l}=j_{J}\cos (\pi \phi _{l}/\phi _{o}).
\end{equation}
Here $j_{c}$ is the lowest junction's critical current in the loop $l$, $%
\phi _{J}$ is the magnetic flux through the junction's area given by 
\begin{equation}
\phi _{J}=BS_{J}=B\left( 2\lambda _{J}+d\right) (r_{J}-2\lambda _{J}),
\end{equation}
and $\phi _{l}$ the magnetic flux through the loops's area taken as 
\begin{equation}
\phi _{l}=BS_{l}=B\pi \left( r_{l}+\lambda _{J}\right) ^{2}.
\end{equation}
At this level we have the fundamental quantities in terms of the basic
granular parameters: the distance between two adjacent grains $d$, the
junction size $r_{J}$, the loop area $r_{l}$ and the magnetic penetration
length $\lambda _{J}$ whose temperature dependence is 
\begin{equation}
\lambda _{J}=\lambda _{o}\left( 1-T/T_{c}\right) ^{-\beta },  \nonumber
\end{equation}
with $\lambda _{o}$ the zero temperature penetration depth and $\beta $ a
positive critical exponent (in BCS theory $\beta =1/2$).

In macroscopic samples, different areas of junctions and loops are possible.
Therefore, the total magnetization current $j_{M}$ can be thought of as the
superposition of a collection of loop currents $j_{l}$ modulated by $j_{J}$.
Thus, to describe the magnetization current we shall consider the expression 
\begin{equation}
j_{M}=\sum_{l,J}j_{c}\left( T,B\right) \left| \frac{\sin \left( \pi \phi
_{J}/\phi _{o}\right) }{\pi \phi _{J}/\phi _{o}}\right| \cos \left( \pi \phi
_{l}/\phi _{o}\right) 
\end{equation}
where 
\begin{equation}
j_{c}\left( T,B\right) =j_{o}\left( 1-T/T_{c}\right) ^{\alpha }\exp \left[ -%
\frac{2\lambda _{o}}{\zeta _{o}}\left( 1-T/T_{c}\right) ^{\alpha -0.5}+\frac{%
2\lambda _{o}}{\zeta _{o}}\right]   \label{suscepy}
\end{equation}
and $\zeta _{o}$ is the zero temperature decoherence length defined as in
reference \cite{pepe1}. In polycrystalline samples many of these parameters
vary randomly. Notice that the temperature dependence is not completely
factorized because both $\phi _{J}$ and $\phi _{l}$ depend also on the
temperature, and $\alpha $ and $\zeta _{o}$ depend on the magnetic field.

On the other hand, taking into account that the magnetization current and
the susceptibility are quantities proportional to each other, it is possible
to conclude that the real part of the susceptibility $\chi $ is described by
a similar function \cite{clem}, i.e. 
\begin{equation}
\mathop{\rm Re}%
\chi _{c}\left( T,B\right) =\chi _{c}\left( 0,B\right) \left(
1-T/T_{c}\right) ^{\alpha }\exp \left[ -\frac{2\lambda _{o}}{\zeta }\left(
1-T/T_{c}\right) ^{\alpha -0.5}+\frac{2\lambda _{o}}{\zeta }\right] 
\end{equation}
The magnetic field dependence is like that in equation (\ref{suscepy}). For
small values of $B$, applied in the experimental measurements discussed
below, we can safely assume that the sum upon the junction and loop indices
can be substituted by a contribution of a junction and a loop with
''effective'' areas $S_{J,eff}$ and $S_{l,eff}$, such that 
\begin{equation}
\chi _{c}\left( 0,B\right) =\chi _{c}\left( 0,0\right) \left| \frac{\sin
\left( \pi BS_{J,eff}/\phi _{o}\right) }{\left( \pi BS_{J,eff}/\phi
_{o}\right) }\right| \cos \left( \pi BS_{l,eff}/\phi _{o}\right)   \nonumber
\end{equation}
These equations will be used, in the next section, to adjust the
experimental points choosing different values for the parameters $\alpha $, $%
\zeta $, $\chi _{o}$, $S_{J,eff}$ and $S_{l,eff}$. Taking $d\approx 0$, and
the ''average'' value of $\lambda _{o}$ ($\lambda _{o}\approx 3000nm$)\cite
{lambda}, it is also possible to estimate the ''effective'' values of $%
r_{J,eff}$ and $r_{l,eff}$.

\section{Experimental results and discussion}

As mentioned above, a more flexible but rigorous application of the
definition of the critical parameter lead us, as in previous works, to
determine reliable and precise measurements of critical transport currents
as functions of the magnetic field, based on direct transport critical
currents measurements for different temperatures, \cite
{ourspb,ourspc1,ourspc2}. Similarly, direct temperature dependence of the
magnetic susceptibility $\chi $ in polycrystalline superconducting samples
allows us to determine, with high precision, the magnetic field dependence
of $\chi $ for different fixed temperatures.

In the experimental procedure, partially explained in references \cite
{ourspb,ourspc1,ourspc2}, the temperature is slowly raised by natural
heating during the experiment, while the external magnetic field is kept
constant. In this way, a set of temperature dependent susceptibility data is
obtained. Changing the magnetic field, a new set of data $\chi =\chi
_{B}\left( T\right) $, with $B$ taken as a parameter, is obtained. All these
data $\chi _{B}\left( T\right) $, corresponding to different applied
magnetic fields, generate a surface in the $B$-$T$-$\chi $ space and define
the susceptibility $\chi =\chi \left( B,T\right) $.

A series of $Y$-based samples were prepared by standard solid state
reactions. The mixture of high purity $Y_{2}O_{3}$, $Ba_{2}CO_{3}$ and $CuO$
powders were ground, pelleted at a pressure of $5tons/in^{2}$ and pre-fired
at $910^{o}C$ for 24 hours. This procedure was repeated three times for
heating temperatures of $920^{o}C$, $930^{o}C$ and $940^{o}C$. An extra
sintering of compact pellets in oxygen flow at $950^{o}C$ for $24h$ was
followed by a slow cooling in oxygen flow. The cooling rate of this stage
was varied in order to get two samples with different grain sizes. In Figure
1 we observe the different structures of samples G and B that will be
analyzed in this paper -- sample B with grains separated by inter granular
dark regions and sample G, with closely packed grains and a minimum spacing
between them (see Figure 1).

The samples were cut and polished in the shape of thin disks and then
mounted on cooper paste in order to decrease the temperature gradient. The
temperature of the sample was measured with a platinum resistor and a Lake
Shore detector. These samples were cooled in an APD SCS cryostat adapted for 
$AC$ susceptibility measurements. The output signal was processed with a
Lock-in amplifier.

In Figure 2, the susceptibilities of samples G and B, are plotted together
to make evident the influence of the physical structure on the behavior of $%
\chi $ as a function of the temperature and the external magnetic field. In
the case of sample B, $\chi $ takes relatively low values even for a low
temperature ($35K$) and zero magnetic field (see Figure 5), and the constant
temperature susceptibility as a function of the magnetic field decreases
with visible oscillations due to the Josephson junction effects. The
susceptibility reduces rapidly with increasing temperature, from $0.49$ (at $%
T=35K$) to $0.38$ (at $T=70K$). At $B\approx 250\mu T$, the magnetic field
effect on $\chi $ is weaker because some junctions are already disconnected
and $\chi $ is controlled mainly by the diamagnetic properties of the
grains. The high-temperature curve oscillates with the magnetic field less
than the low temperature curve for the same reason.

The susceptibility shown in Figure 6 for sample G exhibits a completely
different behavior and higher values ($\approx 0.73$ for $T=35K$ and $%
B\approx 50\mu T$). It is obvious that the intergrain currents can flow
easier through short junctions, and thus screen the external magnetic field
more efficiently. The susceptibility decreases with the magnetic field, but
without visible oscillations. The reduced width of the Josephson junctions
does not permit an appreciable penetration of the magnetic field into them.
The decrease of $\chi $ with temperature corresponds to an increase of the
type III regions, where vortices of the external magnetic field can
penetrate.

\begin{table}[p]
\caption{Parameters used to fit data, in figures 5 and 6 and some estimated
geometrical parameters. We give here the critical temperature $T_{c}$, the
effective grain and loop areas $S_{J,eff}$ and $S_{l,eff}$, the critical
exponent $\protect\alpha$, the decoherence length $\protect\zeta_{o}$ at $%
T=0 $, and the susceptibility $\protect\chi_{o}$ at $B=0$.. We give also the
estimated values of the Josephson junction and loop effective radius $%
r_{J,eff}$, $r_{l,eff}$, for samples B and G.}
\label{tablita}
\begin{center}
\begin{tabular}{ccc}
\hline\hline
Sample & B & G \\ \hline\hline
$T_{c}$ [$K$] & $91.0$ & $91.0$ \\ 
&  &  \\ 
$S_{J,eff}$ [$\mu m^{2}$] & $0.708$ & $0.834$ \\ 
&  &  \\ 
$S_{l,eff}$ [$\mu m^{2}$] & $507.75$ & $---^{\dag }$ \\ 
&  &  \\ 
$r_{J,eff}$ [$\mu m$] & $1.78$ & $1.99$ \\ 
&  &  \\ 
$r_{l,eff}$ [$\mu m$] & $12.713$ & $---^{\dag }$ \\ 
&  &  \\ 
$\zeta_{o}$ [$nm$] & $41-255$ & $35-211$ \\ 
&  &  \\ 
$\alpha $ & $0.52-0.98$ & $0.53-0.83$ \\ 
&  &  \\ 
$\chi_c\left(0,0\right)$ & $0.4819$ & $0.7273$ \\ 
&  &  \\ \hline\hline
\end{tabular}
\end{center}
\par
$^{\dag}$ Within the fitting resolution negligible loop contribution is
found.
\end{table}

In all cases, it was possible to fit the data by choosing the values of $%
r_{J}$, $r_{l}$, $\zeta $, $\alpha $ and $\chi _{c}(0,0)$, as indicated in
Table \ref{tablita}. Besides the sample-dependent (zero field and zero
temperature) scaling susceptibility factor $\chi _{c}(0,0)$, which comprises
the global magnetic field response, the critical exponent $\alpha $, the
decoherence length $\zeta $, and the Josephson junction and loop effective
areas values, correspond quite well to the samples characteristics. In the
case of sample B (see Figure 5) the adjustment is not exact, due to the
irregular shapes of the grains and junctions. The fitting is far more better
for sample G, where the loops of currents do not play an important role (see
Figure 6). The major difference within the parameter values in Table \ref
{tablita} appears in the parameter $r_{l}$, which for sample G is
practically vanishing, while for sample B it has a relatively high value. In
sample B the porous structure forces the Cooper pairs to larger loops around
the intergrain regions V. On the contrary, in sample B such loops are not
necessary because the grains are arranged very closely.

Concerning the critical exponent $\alpha $ appearing in the polynomial and
exponential temperature dependent factors, we notice that their values
remain between $1/2$ and $1$ (increasing with the magnetic field), as was
found already in previous reports, where the order parameter was shown to
decay exponentially because of the intergrain tunneling process. Finally,
the fitting values of the decoherence length $\zeta $, are more or less the
same for both samples and, as expected, grow with the magnetic field.

\section{Conclusions}

In this paper we studied another physical quantity characterizing the
polycrystalline superconductors electromagnetic properties. We show that, as
for the critical transport currents, the Josephson and loop's interfering
currents have, depending on the sample quality, different degrees of
influence on the magnetic field response. The susceptibility of equations
(8) and (9) gives reasonably good predictions on its behavior as a function
of the temperature and the external magnetic field. The oscillations of $%
\chi $, observed experimentally, are well explained on the basis of the
quantum interference theory, the exponential suppression of the order
parameter induced by tunneling processes, and the polynomial temperature
dependence known from the standard theory. \acknowledgements
We acknowledge discussions with Professor H. Simanjuntak and the partial
support of CONACyT of M\'exico under Project No. 350-E9301 and of The Abdus
Salam International Center for Theoretical Physics, Trieste-Italy.

{\Large {\bf Captions}} \newline

\begin{enumerate}
\item  Microscopic images of the samples G and B.

\item  The susceptibility of the good and bad samples differ in magnitude
and have a qualitatively different begavior as functions of the applied
magnetic field. The curves shown here are for $T=35 K$.

\item  The screening currents distributions and the tunneling path in the
two adjacent ceramic grains. In regions I we assume complete Meissner effect;
in regions II flow the screening currents, with a density distribution as
shown below, reaching a critical value at the boundary of this and regions
III, where the external magnetic field penetrates; finally, the intergrain
region IV is also shown.

\item  An idealized picture of three adjacent grains with Josephson's
junctions between them, forming an elementary tunneling circuit. Each grain
has its own regions of the types I, II, and III, as explained in Figure 3.
The screening currents circulate within regions II in the counter-clock
direction. The intergrain tunneling current 1 circulates in the same
direcion in the outer part of the circuit, while the current 2 in the inner
part of the circuit has the opposite, clockwise direction. The external
magnetic field can penetrate into the region III of each grain, and into the
regions IV and V between the grains.

\item  Behavior of $\chi$ as a function of the applied magnetic field at
different temperatures for the sample B.

\item  Behavior of $\chi$ as a function of the applied magnetic field at
different temperatures for sample G.
\end{enumerate}

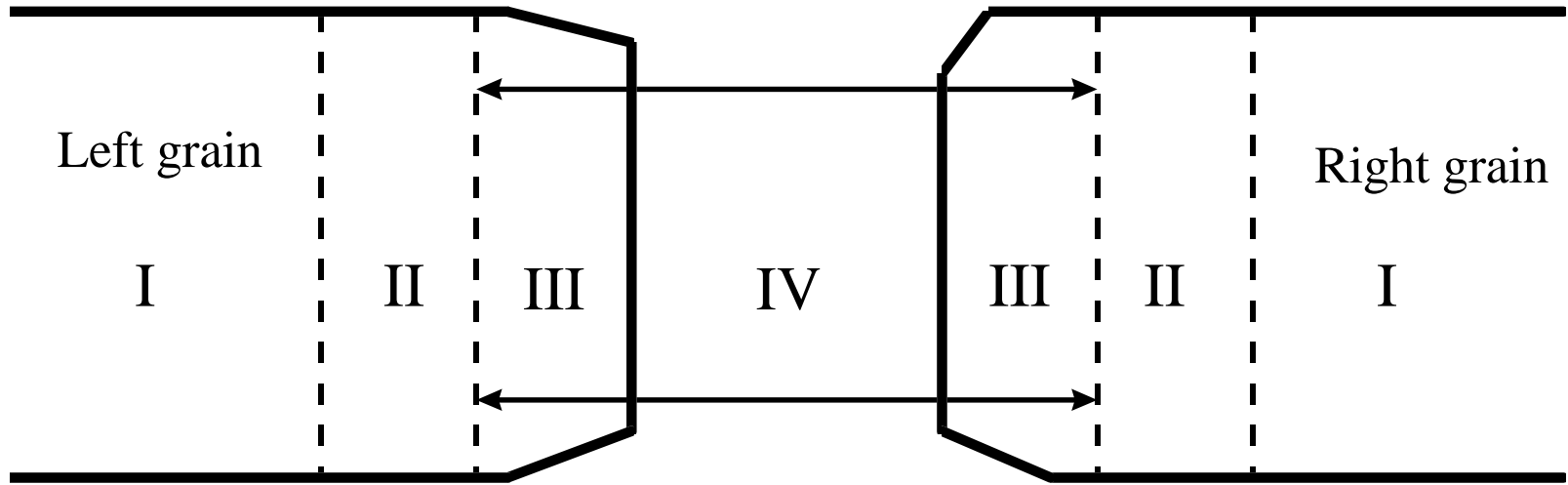
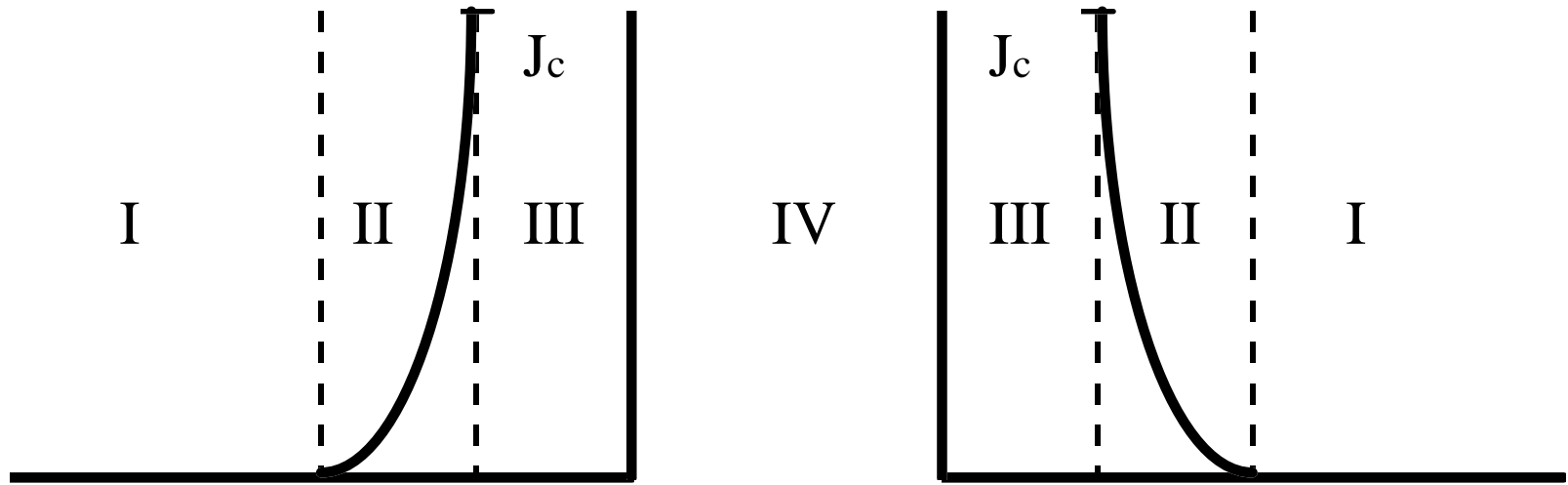
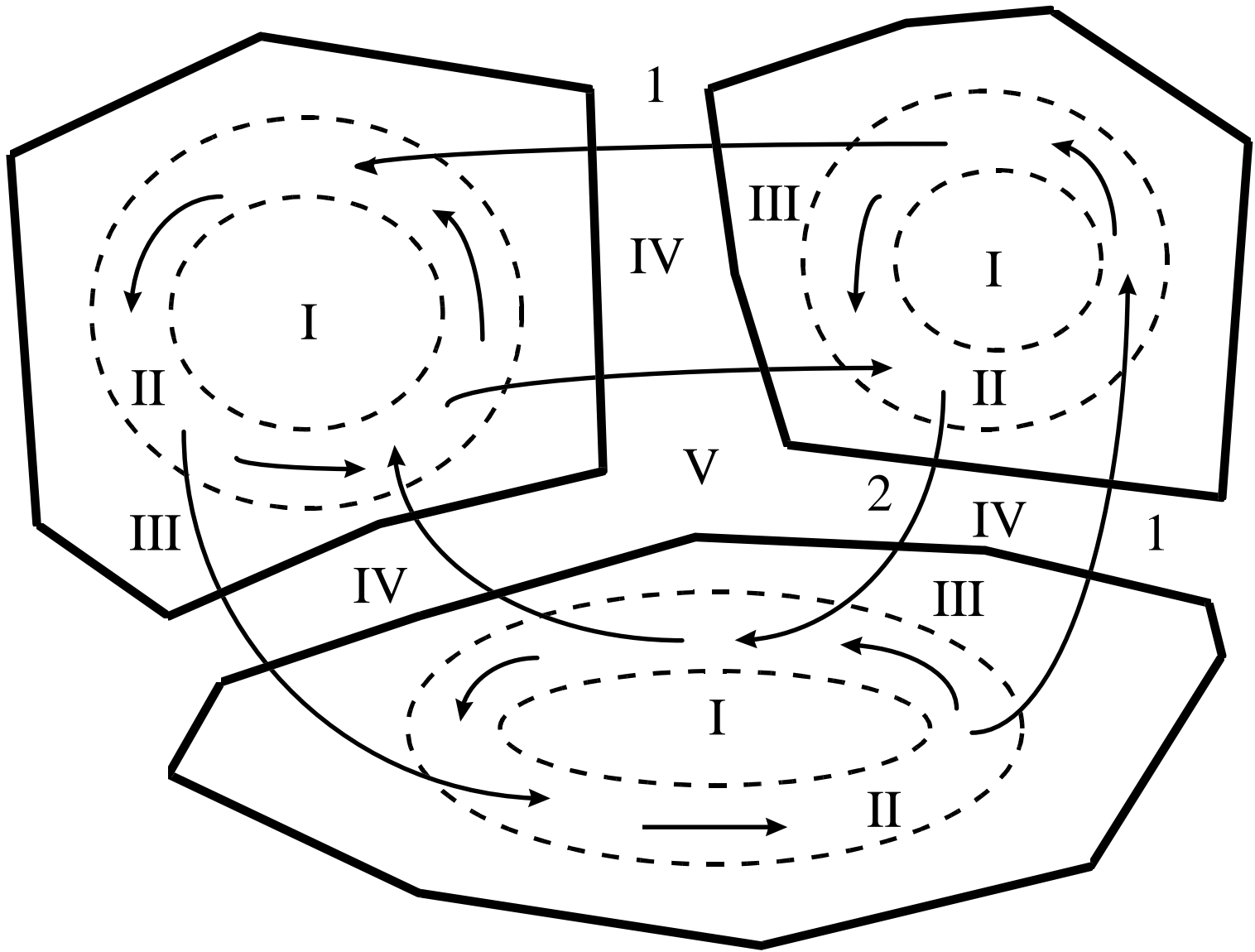
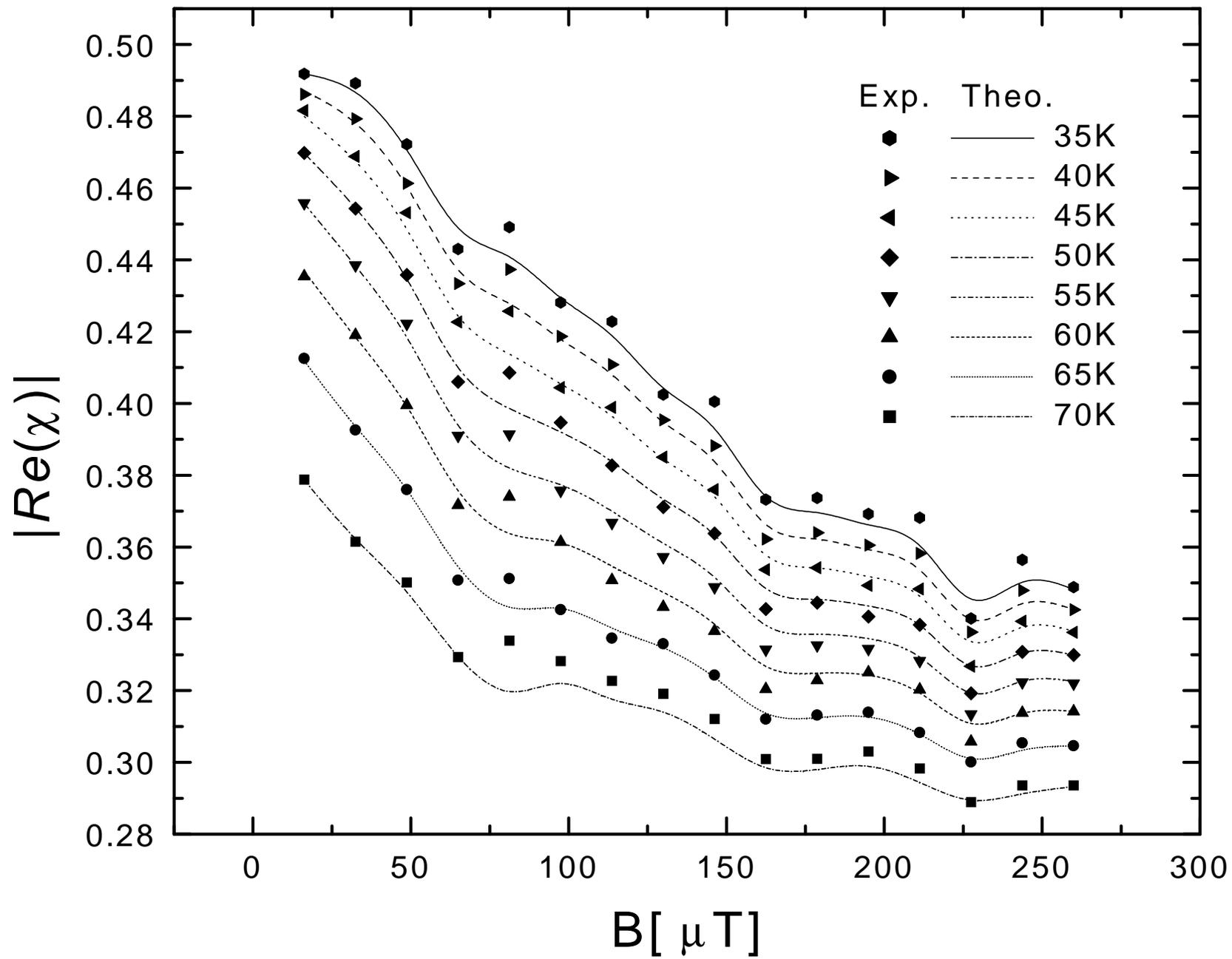
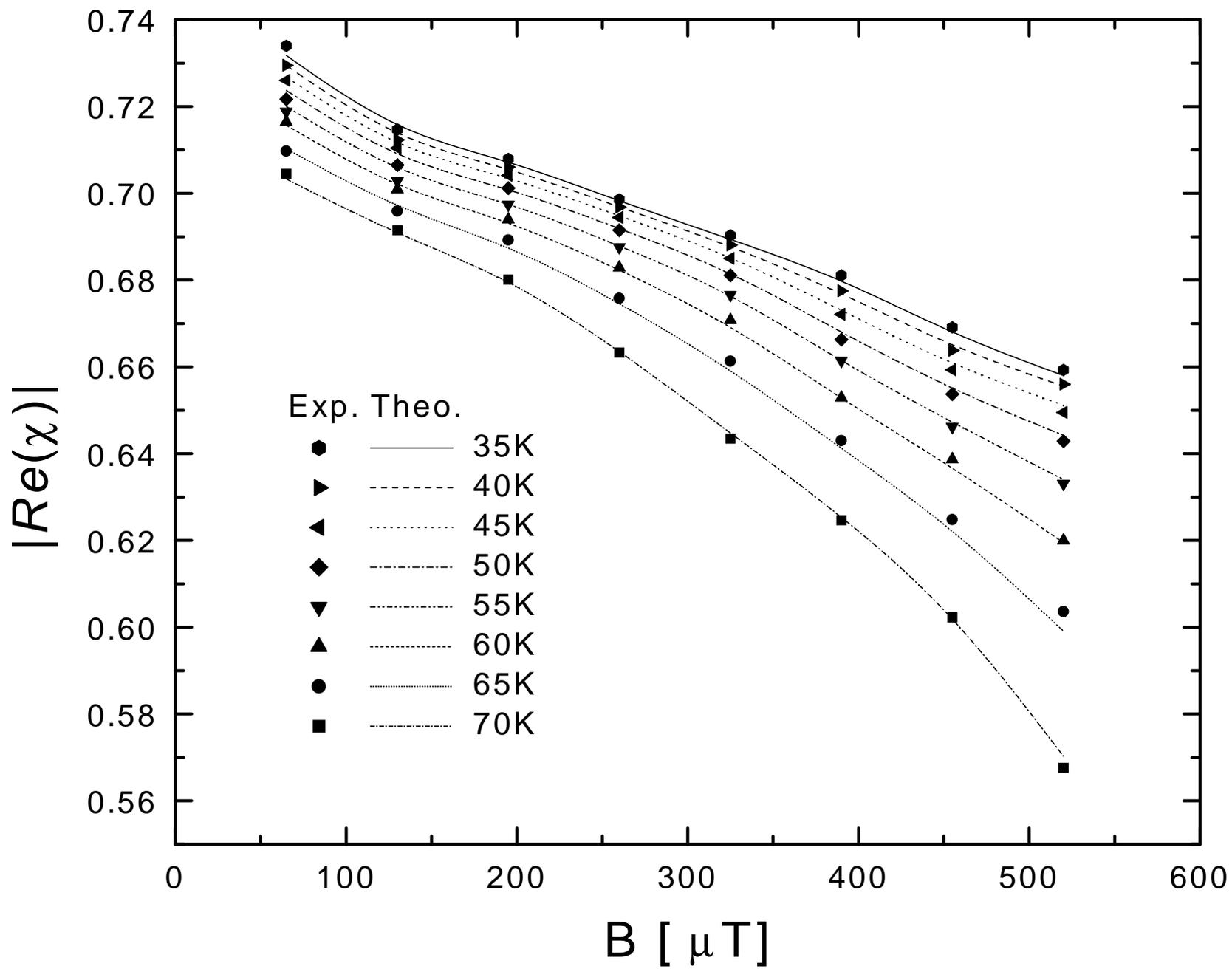
\end{document}